\documentclass[pdftex,twocolumn,epjc3]{svjour3}
\usepackage{amsfonts}
\usepackage{amsmath}
\usepackage{amssymb,bbold}
\usepackage[normalem]{ulem}

\RequirePackage[T1]{fontenc}
\smartqed
\RequirePackage{graphicx}
\RequirePackage{mathptmx}
\RequirePackage{flushend}
\RequirePackage[numbers,sort&compress]{natbib}
\RequirePackage[colorlinks,citecolor=blue,urlcolor=blue,linkcolor=blue]{hyperref}
\journalname{Eur. Phys. J. C}
\RequirePackage{color}

\begin{document}

\title{On planar quantum dynamics of a magnetic dipole moment in the
presence of electric and magnetic fields}
\author{Edilberto O. Silva \thanksref{e1,addr1}}
\thankstext{e1}{e-mail: edilbertoo@gmail.com}
\institute{Departamento de F\'{i}sica,
           Universidade Federal do Maranh\~{a}o,
           Campus Universit\'{a}rio do Bacanga,
           65085-580 S\~{a}o Lu\'{i}s-MA, Brazil\label{addr1}
         }
\date{Received: date / Accepted: date}
\maketitle

\begin{abstract}
The planar quantum dynamics of a neutral particle with a magnetic dipole
moment in the presence of electric and magnetic fields is considered. The
criteria to establish the planar dynamics reveal that the resulting
nonrelativistic Hamiltonian has a simplified expression without making
approximations, and some terms have crucial importance for system dynamics.
\end{abstract}

\section{Introduction}

\label{sec:introduction}

The Aharonov-Bohm effect \cite{PR.1959.115.485} has been an usual framework
\ for demonstrating the importance of potentials in quantum mechanics. After
its experimental verification \cite{PRL.1960.5.3,PRL.1995.74.2847} several
other analogs effects were being proposed along the last decades. For
example, in Ref. \cite{PRL.1984.53.319}, it was shown that a particle with a
magnetic moment moving in an electric field acquires a quantum phase. This
phase has been observed in a neutron interferometer \cite{PRL.1989.63.380}
and in a neutral atomic Ramsey interferometer \cite{PRL.1993.71.3641}. In
Refs. \cite{PRA.1993.47.3424,PRL.1994.72.5} it was verified that a neutral
particle with an electric dipole moment which moves in a magnetic field
acquires a topological phase. The experimental confirmation of this phase
was established in Ref. \cite{PRL.1995.75.2071}. In Ref. \cite%
{PRL.2000.85.1354} was proposed a unified and fully relativistic treatment
of the interaction of the electric and magnetic dipole moments of a particle
with the electromagnetic field. The essence of this study reveals that new
force on dipoles are obtained using the non-Abelian nature of this
interaction, and new experiments analogous to the Aharonov-Bohm effect to
test this interaction are proposed. Since this interaction is a consequence
of a nonminimal coupling is also interesting to analyze the consequences of
this interaction in other contexts. For instance, it may be of interest to
study scattering and bound states of neutral fermions in external
electromagnetic fields and access other physical quantities such as energy
bound states \cite{JPA.1993.26.5631,EPJC.2013.73.2402} and scattering \cite%
{PRA.2010.81.12710,PRA.2005.72.042103}. The dual results of the magnetic
dipole interaction for the electric dipole interaction for the system
considered in Ref. \cite{PRL.2000.85.1354} has been established in Ref. \cite%
{PLA.1989.138.347}, where the phase shift in the interference of a magnetic
or electric dipole due to the electromagnetic field is obtained
relativistically and non-relativistically.

In this work, we consider the same system addressed in Ref. \cite{PLA.1989.138.347} but now assuming that the dynamics is purely planar, and
derive its equation of motion. This system is a generalization, for example, from that studied in Ref. \cite{EPJC.2013.73.2402}, where only the effects of an electric field has been considered. An interesting feature of this system is that even in the absence of electric field it admits bound and scattering states, which does not occur in the previous work. As an application, we consider the problem of
bound state for the case of a magnetic dipole moment interacting with
electric and magnetic fields generated by an infinitely long charged
solenoid, carrying a magnetic field. In our treatment, we consider the
self-adjoint extension method \cite{Book.1975.Reed.II}, which is appropriate to address any system
endowed with a singular Hamiltonian (due to localized fields sources or
quantum confinement) \cite{Book.1988.Demkov,Book.1988.Albeverio,PLB.2013.719.467,PRD.1989.40.1346,LMP.1998.43.43,AoP.2012.327.2742,EPJC.2013.73.2548,EPJC.2014.74.2708,TMP.2010.163.511,MPLA.2012.27.1250027}. We determine the energy spectrum and wave functions by
applying boundary conditions allowed by the system.

\section{The planar Pauli equation}

We begin with the Dirac equation in $(3+1)$ dimensions \cite%
{PLA.1989.138.347} which governs the nonrelativistic dynamics of a neutral
particle that possesses magnetic dipole moment, in the presence of electric
and magnetic fields $\left( \hbar =c=1\right) $
\begin{equation}
\biggl[i\gamma ^{\mu }\partial _{\mu }-M+\frac{\mu }{2}\sigma ^{\mu \nu
}F_{\mu \nu }\biggr]\Psi =0,  \label{eqcd}
\end{equation}%
where $\mu $ is the magnetic dipole moment, $F_{\mu \nu }$ is the
electromagnetic tensor whose components are given by $\left(
F^{0i},F^{ij}\right) =\left( -E^{i},-\varepsilon ^{ijk}B^{k}\right) $, and $%
\left( \sigma ^{0j},\sigma ^{ij}\right) =\left( i\alpha ^{j},-\epsilon
_{ijk}\Sigma ^{k}\right) $, where $\Sigma ^{k}$ is the spin operator, are
the components of the operator $\sigma ^{\mu \nu }=i[\gamma ^{\mu },\gamma
^{\nu }]/2$, which are given in terms of the Dirac matrices. \ With this
notation, it is possible to show that the spin is coupled to the
electromagnetic field tensor through the term
\begin{equation}
\frac{1}{2}\sigma ^{\mu \nu }F_{\mu \nu }=-\pmb{\Sigma }\cdot \mathbf{%
B}+i\pmb{\alpha }\cdot \mathbf{E}
\end{equation}%
where $\mathbf{E}$ and $\mathbf{B}$ are the electric and magnetic field
strengths. This result is explicitly calculated in the following
representation of the $\gamma $-matrices:
\begin{align}
\gamma ^{0}& =\left(
\begin{array}{cc}
1 & 0 \\
0 & -1%
\end{array}%
\right) ,\;\;\; \pmb{\gamma}=\left(
\begin{array}{cc}
0 & \pmb{\sigma} \\
-\pmb{\sigma} & 0%
\end{array}%
\right) ,  \notag \\
\pmb{\alpha}& =\gamma ^{0}\pmb{\gamma}=\left(
\begin{array}{cc}
0 & \pmb{\sigma} \\
\pmb{\sigma} & 0%
\end{array}%
\right), \;\;\;\pmb{\Sigma}=\left(
\begin{array}{cc}
\pmb{\sigma} & 0 \\
0 & \pmb{\sigma}%
\end{array}%
\right) .  \notag
\end{align}
with $\pmb{\sigma }=(\sigma _{1},\sigma _{1},\sigma _{3})$ being the
Pauli matrices. Equation (\ref{eqcd}) can be written as%
\begin{equation}
\hat{H}_{D}\Psi =\mathcal{E}\Psi ,  \label{eqdh}
\end{equation}%
where the operator%
\begin{equation}
\hat{H}_{D}=\beta M+\pmb{\alpha }\cdot \mathbf{p}+\mu (-\beta \pmb{%
\Sigma }\cdot \mathbf{B}+i\pmb{\gamma }\cdot \mathbf{E})  \label{hD}
\end{equation}%
is the Dirac Hamiltonian. The non-relativistic limit of Eq. (\ref{eqdh}) was
established in Ref. \cite{PLA.1989.138.347}, and the relevant equation is
found to be:
\begin{equation}
\hat{H}\psi =E\psi ,  \label{spe}
\end{equation}%
where $\psi $ is a two-component spinor, with
\begin{equation}
\hat{H}=\frac{1}{2M}\left[ \mathbf{p}-\left( \pmb{\mu }\times \mathbf{E}%
\right) \right] ^{2}-\frac{1}{2M}\mu ^{2}E^{2}+\frac{1}{2M}\mu \left(
\pmb{\nabla }\cdot \mathbf{E}\right) -\left( \pmb{\mu }\cdot \mathbf{B}%
\right) ,  \label{hm}
\end{equation}%
where $\pmb{\mu }=\mu \pmb{\sigma }$. Our goal is to analyze the physical implications of the Hamiltonian (\ref%
{hm}), when we assume that the system dynamics is now planar. This is
established as follows. By detaching the third component of Eq. (\ref{hm}),
we get
\begin{eqnarray}
\hat{H} &=&\frac{1}{2M}\sum_{i=1}^{2}\left[ \mathbf{p}_{i}-\left(
\pmb{\mu }\times \mathbf{E}\right) _{i}\right] ^{2}-\frac{1}{2M}\mu
^{2}E^{2} \notag \\
&+&\frac{1}{2M}\left[ \mathbf{p}_{3}-\left( \pmb{\mu }\times \mathbf{E}%
\right) _{3}\right] ^{2}  \notag \\
&+&\frac{1}{2M}\sum_{i=1}^{2}\mu \left( \pmb{\nabla }\cdot
\mathbf{E}\right) _{i}+\frac{1}{2M}\mu \left( \pmb{\nabla }\cdot \mathbf{E%
}\right) _{3},  \notag \\
&-&\sum_{i=1}^{2}\left( \pmb{\mu }\cdot \mathbf{B}\right)
_{i}-\left( \pmb{\mu }\cdot \mathbf{B}\right) _{3},~~~(i=1,2).
\label{hdec}
\end{eqnarray}%
If we assume that dynamic is planar, the above Hamiltonian provides an
important result, namely, the $\left[\mathbf{p}_{3}-\left( \pmb{\mu }%
\times \mathbf{E}\right) _{3}\right] ^{2}$ term leads exactly to the
quantity $\mu ^{2}E^{2}/2M$. The planar case is accessed by requiring that $%
p_{z}=z=0$ together with the imposition of the fields should not have the
third direction. This question can also be understood when we look at the
symmetry under $z$ translations, which allows us to access the solutions of
the planar Dirac equation. This type of simplification is in fact manifested
only when we assume that the particle moves in the plane. Thus, since $%
A^{\mu }=\left( \Phi ,\mathbf{A}\right) $, we write the electric and
magnetic fields, respectively, as%
\begin{eqnarray}
\mathbf{E} &=&-\mathbf{\hat{x}\partial }_{x}\Phi \left( x,y\right) -\mathbf{%
\hat{y}\partial }_{y}\Phi \left( x,y\right) ,  \label{cEr} \\
\mathbf{B} &=&\mathbf{\hat{z}}\left( \partial _{x}A_{y}-\partial
_{y}A_{x}\right) ,  \label{cB}
\end{eqnarray}%
Fields $\mathbf{E}$ and $\mathbf{B}$ above are now intrinsically
two-dimensional. Note that the square of Eq. (\ref{cEr}) gives exactly $%
E_{1}^{2}+E_{2}^{2}$ for the planar case. Also, the restriction imposed on
the potential $\mathbf{A}$ reveals that the $\left( \pmb{\mu }\cdot
\mathbf{B}\right) _{i}$ term in Eq. (\ref{hdec}) is now identically zero.
Now, we can show that the quantity $\left( \pmb{\mu }\times \mathbf{E}%
\right) _{3}$ is given by $\mu \left( \sigma _{1}E_{2}-\sigma
_{2}E_{1}\right) $, and the third term of Eq. (\ref{hdec}) results
\begin{equation}
\left[ \mathbf{p}_{3}-\left( \pmb{\mu }\times \mathbf{E}\right) _{3}%
\right] ^{2}\rightarrow \mu ^{2}\left( \sigma _{1}E_{2}-\sigma
_{2}E_{1}\right) ^{2}=\mu ^{2}E^{2}.  \label{redE}
\end{equation}%
Thus, we now can write Eq. (\ref{hdec}) as
\begin{equation}
\hat{H}=\frac{1}{2M}\sum_{i=1}^{2}\left\{ \left[ \mathbf{p}%
_{i}-\left( \pmb{\mu }\times \mathbf{E}\right) _{i}\right] ^{2}+\mu
~\left( \pmb{\nabla }\cdot \mathbf{E}\right) _{i}\right\} -\left( \pmb{%
\mu }\cdot \mathbf{B}\right) _{3},  \label{hE}
\end{equation}%
where the magnetic interaction term $\left( \pmb{\mu }\cdot \mathbf{B}%
\right) _{3}$ gives the only explicit dependence of the spin. In Ref. \cite%
{PLA.1989.138.347} it was assumed that the charge density $\rho =\pmb{\nabla }\cdot \mathbf{E}$ and also the $\mu ^{2}E^{2}/2M$ term (for thermal
neutrons) are negligible. In fact, this approximation can only be performed,
if we are only interested in the study of phase shift. However, when we want
to study the dynamics of the system, such as the scattering and bound states
problems, all terms of the equation motion of must be taken into account.
This, for example, has been addressed by Hagen \cite{PRL.1990.64.2347} to
show that there is an exact equivalence between the AB and AC effects for
spin-$1/2$ particles. In these effects, the $\pmb{\nabla }\cdot \mathbf{E}
$ and $\pmb{\nabla }\times \mathbf{A}$ being proportional to a delta
function, such terms must now contribute to the dynamics of the system, and
can not be neglected. For this reason, since we are dealing with
Aharonov-Bohm-like system for spin-1/2 particles, the $\mu \left( \pmb{%
\nabla }\cdot \mathbf{E}\right) _{i}$ and $\left( \pmb{\mu }\cdot \mathbf{%
B}\right) _{3}$ terms in Eq. (\ref{hE}), may not be negligible. Let's
clarify this issue. Consider an infinitely long solenoid, carrying a
magnetic field $\mathbf{B}$, and with a charge density $\lambda $
distributed uniformly about it along the $z$-axis. The electric field and
magnetic flux tube (in cylindrical coordinates) generated by this
configuration are known to be
\begin{eqnarray}
\mathbf{E} &=&2\lambda \frac{\mathbf{\hat{r}}}{r},~~~\pmb{\nabla }%
\cdot \mathbf{E}=2\lambda \frac{\delta (r)}{r},  \label{vtE} \\
\mathbf{B} &=&\pmb{\nabla }\times \mathbf{A}=\phi \frac{\delta (r)}{r}%
\mathbf{\hat{z}},  \label{vtB}
\end{eqnarray}%
where $\phi $ is the magnetic flux inside the tube, and the vector potential
in the Coulomb gauge is
\begin{equation}
\mathbf{A}=\frac{\phi }{r}\pmb{\hat{{\varphi }}},  \label{vectora}
\end{equation}%
We see that the fields $\mathbf{E}$ and $\mathbf{B}$ are proportional to a $%
\delta $ function. By using Eqs. (\ref{vtE}) and (\ref{vtB}), Pauli equation
(\ref{spe}) is now written as%
\begin{equation}
\hat{H}\psi =E\psi ,  \label{Paulif}
\end{equation}%
with%
\begin{equation}
\hat{H}=\frac{1}{2M}\sum_{i=1}^{2}\left( \mathbf{p}_{i}-\phi
_{E}\sigma _{z}\frac{\pmb{\hat{\varphi}}}{r}\right) ^{2}+\left( \phi
_{E}-\phi _{B}\sigma _{z}~\right) \frac{\delta (r)}{r}.  \label{Hmf}
\end{equation}%
where $\phi _{E}=2\mu \lambda $ and $\phi _{B}=2M\mu \phi $. From Eq. (\ref%
{Paulif}), we can see that $\psi $ is an eigenfunction of $\sigma _{z}$,
whose eigenvalues designate by $s=\pm 1$, that is, $\sigma _{z}\psi =\pm
\psi =s\psi $. Thus, since $\hat{H}$ commutes with the operators $\hat{J}%
_{z}=-i\partial _{\varphi }+\sigma _{z}/2$, where $\hat{J}_{z}$ is the total
angular momentum operator in the $z$-direction, we seek solutions of the
form
\begin{equation}
\psi (r,\varphi )=\left[
\begin{array}{c}
f_{m}(r)\;e^{im\varphi } \\
g_{m}(r)\;e^{i(m+1)\varphi }%
\end{array}%
\right] ,  \label{wavef}
\end{equation}%
with $m+1/2=\pm 1/2,\pm 3/2,\ldots ,$ $(m\in \mathbb{Z)}$. Inserting (\ref%
{wavef}) into Eq. (\ref{Paulif}), we can extract the radial equation for $%
f_{m}(r)$
\begin{equation}
Hf_{m}(r)=k^{2}f_{m}(r),  \label{eigen}
\end{equation}%
where
\begin{equation}
H=H_{0}+\left( \phi _{E}-s\phi _{B}\right) \frac{\delta (r)}{r},
\label{hfull}
\end{equation}%
and
\begin{equation}
H_{0}=-\frac{d^{2}}{dr^{2}}-\frac{1}{r}\frac{d}{dr}+\frac{\left( m-s\phi
_{E}\right) ^{2}}{r^{2}}.  \label{hzero}
\end{equation}
Note that, even in the absence of electric field, bound and scattering states are possible. This does not occur, for example, in the system studied in Ref. \cite{EPJC.2013.73.2402}, where the particle interacts only with an electric field. Moreover, if a magnetic field is present, the physical system changes completely. We will see later that this fact directly influences on the expression for the self-adjoint extension parameter and, hence, on the boundary conditions allowed by the operator $H_ {0}$. In other words, this has direct implications in the dynamics of the system. This can be seen more easily by studying the signal of $\phi _{E}-s\phi _{B}$ in Eq. (\ref{hfull}), where several possible combinations of $\phi _{E}$, $\phi _{B}$ and $s$ gives us the possibilities for the existence of bound and scattering states. As a result of these combinations, we have
\begin{eqnarray}
\phi _{E}-s\phi _{B} &<&0,~~~\text{scattering~and~bound~states}, \\
\phi _{E}-s\phi _{B} &>&0,~\ ~\text{scattering~states}.
\end{eqnarray}%
The case $\phi _{E}=s\phi _{B}$ is not of interest here because it cancels
the term that explicitly depends on the spin.

\section{Physical regularization and the bound states problem}

In this section, we study the dynamics of the system in all space, including
the $r=0$\ region. We consider the problem of bound states. To this end, we
use the self-adjoint extension method in the treatment. As is well known, if
the Hamiltonian has a singularity point, as is the case of the Hamiltonian
in Eq. (\ref{hfull}), we must verify that it is self-adjoint in the region
of interest. Even though $H_{0}^{\dagger }=H_{0}$, their domains could be
different. This is the crucial point in our study. The operator $H_{0}$,
with domain $\mathcal{D}(H_{0})$, is self-adjoint if $\mathcal{D}%
(H_{0}^{\dagger })=\mathcal{D}(H_{0})$ and $H_{0}^{\dagger }=H_{0}$.
However, for this to be established, we must find the deficiency subspaces,%
\begin{eqnarray}
N_{+} &=&\left\{ \psi \in \mathcal{D}(H_{0}^{\dagger }),H_{0}^{\dagger }\psi
=z_{+}\psi ,\mbox{Im}\,z_{+}>0\right\} , \\
N_{-} &=&\{\psi \in \mathcal{D}(H_{0}^{\dagger }),H_{0}^{\dagger }\psi
=z_{-}\psi ,\mbox{Im}\,z_{-}<0\},
\end{eqnarray}
with dimensions $n_{+}$ and $n_{-}$, respectively, called deficiency indices
of $H_{0}$ \cite{Book.1975.Reed.II}. We also know of this theory that a
necessary and sufficient condition for $H_{0}$ being essentially
self-adjoint is that its deficiency indices $n_{+}=n_{-}=0$. On the other
hand, if $n_{+}=n_{-}\geq 1$ the operator $H_{0}$ has an infinite number of
self-adjoint extensions parametrized by the unitary operators $%
U:N_{+}\rightarrow N_{-}$. With these ideas in mind, we now decompose the
Hilbert space $\mathcal{H}=L^{2}(\mathbb{R}^{2})$ with respect to the
angular momentum $\mathcal{H}=\mathcal{H}_{r}\otimes \mathcal{H}_{\varphi }$%
, where $\mathcal{H}_{r}=L^{2}(\mathbb{R}^{+},rdr)$ and $\mathcal{H}%
_{\varphi }=L^{2}(S^{1},d\varphi )$, with $S^{1}$ denoting the unit sphere
in $\mathbb{R}^{2}$. The operator $-{\partial _{\varphi }^{2}}$ is known to
be essentially self-adjoint in $L^{2}(S^{1},d\varphi )$. By using the
unitary operator \cite{Book.1988.Albeverio}
\begin{equation}
V:L^{2}(\mathbb{R}^{+},rdr)\rightarrow L^{2}(\mathbb{R}^{+},dr),
\end{equation}%
given by
\begin{equation}
(VQ)(r)=r^{1/2}Q(r),
\end{equation}%
the operator $H_{0}$ reads
\begin{equation}
H_{0}^{^{\prime }}=VH_{0}V^{-1}=-\frac{1}{2M}\left\{ \frac{d^{2}}{dr^{2}}+%
\frac{1}{r^{2}}\left[ \left( m-s\phi _{E}\right) ^{2}-\frac{1}{4}\right]
\right\} ,
\end{equation}%
which is essentially self-adjoint for $\left( m-s\phi _{E}\right) \geq 1$,
while for $\left( m-s\phi _{E}\right) <1$, it admits a one-parameter family
of self-adjoint extensions \cite{Book.1975.Reed.II}. To characterize this
family, we follow the recipe based in boundary conditions given in Ref. \cite%
{CMP.1991.139.103}. Basically, the boundary condition is a match of the
logarithmic derivatives of the zero-energy solutions for Eq. (\ref{eigen})
and the solutions for the problem $H_{0}$ plus self-adjoint extension. Then,
following \cite{CMP.1991.139.103}, we temporarily forget the $\delta $%
-function potential and find the boundary conditions allowed for $H_{0}$.
Next, we substitute the problem in Eq. (\ref{eigen}) by
\begin{equation}
H_{0}f_{\zeta }\left( r\right) =k^{2}f_{\zeta }\left( r\right) ,
\label{ideal}
\end{equation}%
plus self-adjoint extensions. Here, $f_{\zeta }$ is labeled by the parameter
$\zeta $ of the self-adjoint extension, which is related to the behavior of
the wave function at the origin. In order for the $H_{0}$ to be a
self-adjoint operator in $\mathcal{H}_{r}$, its domain of definition has to
be extended by the deficiency subspace, which is spanned by the solutions of
the eigenvalue equation
\begin{equation}
H_{0}^{\dagger }f_{\pm }\left( r\right) =\pm ik_{0}^{2}f_{\pm }\left(
r\right) ,  \label{eigendefs}
\end{equation}%
where $k_{0}^{2}\in \mathbb{R}$ is introduced for dimensional reasons. Since
$H_{0}^{\dagger }=H_{0}$, the only square integrable functions which are
solutions of Eq. (\ref{eigendefs}) are the modified Bessel functions of
second kind
\begin{equation}
f_{\pm }\left( r\right) =K_{m-s\phi _{E}}(\sqrt{\mp i}k_{0}r),  \label{fmm}
\end{equation}%
with $\mbox{Im}\sqrt{\pm i}>0$. By studying Eq. (\ref{fmm}), we verify that
it is square integrable only in the range $m-s\phi _{E}\in (-1,1)$. In this
interval, nevertheless, the Hamiltonian (\ref{hzero}) is not self-adjoint.
The dimension of such deficiency subspace is $(n_{+},n_{-})=(1,1)$. So, we
have two situations for $m-s\phi _{E}$, i.e.,
\begin{align}
-1& <m-s\phi _{E}<0,  \notag \\
&  \label{jrange} \\
0& <m-s\phi _{E}<1.  \notag
\end{align}%
To address both cases of Eq. (\ref{jrange}), we write Eq. (\ref{fmm}) as
\begin{equation}
f_{\pm }\left( r\right) =K_{|m-s\phi _{E}|}(\sqrt{\mp i}k_{0}r).
\label{fmmn}
\end{equation}%
Equation (\ref{fmmn}) allows us to identify the domain of $H_{0}^{\dagger }$
as
\begin{equation}
\mathcal{D}(H_{0}^{\dag })=\mathcal{D}(H_{0})\oplus N_{+}\oplus N_{-}.
\label{dom}
\end{equation}%
So, to extend the domain $\mathcal{D}(H_{0})$ and make it equal to $\mathcal{%
D}(H_{0}^{\dag })$ and therefore to turn $H_{0}$ self-adjoint, we get
\begin{equation}
\mathcal{D}(H_{0,\varsigma })=\mathcal{D}(H_{0}^{\dag })=\mathcal{D}%
(H_{0})\oplus N_{+}\oplus N_{-}.  \label{dominio}
\end{equation}%
Equation (\ref{dominio}) establishes the following result. For each value of
$\zeta $, we have a possible domain for $\mathcal{D}(H_{0,\zeta })$, but are
the physical parameters of the problem that will select a particular value
of it. The Hilbert space is now specified by \cite{Book.1975.Reed.II}
\begin{eqnarray}
f_{\zeta }(r) &=&f_{m}(r)  \notag \\
&+&C\left[ K_{|m-s\phi _{E}|}(\sqrt{-i}k_{0}r)+e^{i\zeta }K_{|m-s\phi _{E}|}(%
\sqrt{i}k_{0}r)\right] ,  \label{domain}
\end{eqnarray}%
where $f_{m}(r)$, with $f_{m}(0)=\dot{f}_{m}(0)=0$ ($\dot{f}\equiv df/dr$),
is the regular wave function and the parameter $\zeta \in \lbrack 0,2\pi )$
represents a choice for the boundary condition. For each $\zeta $, we have a
possible domain for $H_{0}$ and the physical situation is the factor that
will determine the value of $\zeta $ \cite%
{AoP.2010.325.2529,AoP.2008.323.3150,PRD.1989.40.1346,JMP.2012.53.122106,AoP.2013.339.510,PRD.2012.85.041701,EPL.2013.101.51005}%
. Thus, to find a fitting for $\zeta $ compatible with the physical
situation, a physically motivated form for the magnetic field is preferable
for the regularization of the $\delta $-function. This is accomplished by
replacing (\ref{vectora}) with \cite{PRL.1990.64.503}
\begin{equation}
e\mathbf{A}=\left\{
\begin{array}{lr}
\displaystyle\frac{\phi }{r}\pmb{\hat{\varphi}}, & r>a \\
&  \\
0, & r<a.%
\end{array}%
\right.
\end{equation}%
With this modification, the delta function in Eq. (\ref{hfull}) is now
regularized as $\delta (r-a)/a$. A remarkable feature of this regularization
is that, although the functional structure of $\delta (r)/r$ and $\delta
(r-a)/a$ are quite different, we are free to use any form of potential once
that the specific details of the regularization model can be shown to be
irrelevant provided that only the contribution is independent of angle and
has no $\delta $-function contribution at the origin \cite{PRL.1990.64.503}.
It should also be mentioned that the $\delta (r-a)/a$ potential is one
dimensional and well defined contrary to the two dimensional $\delta (r)/r$.

Now, we are in the position to determine a fitting value for $\zeta $. To do
so, we consider the zero-energy solutions for $f_{0}\left( r\right) $ with
the regularization, and for $f_{\zeta ,0}\left( r\right) $ without the $%
\delta $ function, respectively, i.e.,
\begin{eqnarray}
\Bigg \{-\frac{1}{r}\frac{d}{dr}\left( r\frac{d}{dr}\right)  &+&\frac{\left(
m-s\phi _{E}\right) ^{2}}{r^{2}}  \notag \\
&+&\left( \phi _{E}-s\phi _{B}\right) \frac{\delta (r-a)}{a}\Bigg \}%
f_{0}\left( r\right) =0,  \label{statictrue}
\end{eqnarray}%
\begin{equation}
\left\{ -\frac{1}{r}\frac{d}{dr}\left( r\frac{d}{dr}\right) +\frac{\left(
m-s\phi _{E}\right) ^{2}}{r^{2}}\right\} f_{\zeta ,0}\left( r\right) =0.
\label{rhostatic}
\end{equation}%
The value of $\zeta $ is determined by the boundary condition
\begin{equation}
\lim_{a\rightarrow 0^{+}}a\frac{\dot{f}_{0}\left( r\right) }{f_{0}\left(
r\right) }\Big|_{r=a}=\lim_{a\rightarrow 0^{+}}a\frac{\dot{f}_{\zeta
,0}\left( r\right) }{f_{\zeta ,0}\left( r\right) }\Big|_{r=a}.
\label{logder}
\end{equation}%
By integrating Eq. (\ref{statictrue}) from $0$ to $a$ and noting that the
behavior of $f_{0}$ as $a\rightarrow 0$ is $f_{0}\sim r^{\left\vert m-s\phi
_{E}\right\vert }$, the left-hand side of Eq. (\ref{logder}) is found to be
\begin{equation}
\lim_{a\rightarrow 0^{+}}a\frac{\dot{f}_{0}\left( r\right) }{f_{0}\left(
r\right) }\Big|_{r=a}=\phi _{E}-s\phi _{B}.  \label{nrs}
\end{equation}%
To calculate the right-hand side of Eq. (\ref{logder}), we need to use the
asymptotic behavior for $K_{\nu }(z)$ in the limit $z\rightarrow 0$, given
by
\begin{equation}
K_{\nu }(z)\sim \frac{\pi }{2\sin (\pi \nu )}\left[ \frac{z^{-\nu }}{2^{-\nu
}\Gamma (1-\nu )}-\frac{z^{\nu }}{2^{\nu }\Gamma (1+\nu )}\right] .
\label{besselasympt}
\end{equation}%
The substitution of Eq. (\ref{besselasympt}) into Eq. (\ref{domain}) leads
to
\begin{equation}
\lim_{a\rightarrow 0^{+}}a\frac{\dot{f}_{\zeta ,0}\left( r\right) }{f_{\zeta
,0}\left( r\right) }\Big|_{r=a}=\lim_{a\rightarrow 0^{+}}\frac{\dot{W}%
_{\zeta }(r)}{W_{\zeta }(r)}\Big|_{r=a},  \label{dright}
\end{equation}%
with
\begin{eqnarray}
W_{\zeta }(r) &=&\left[ \frac{\left( \sqrt{-i}k_{0}r\right) ^{-\left\vert
m-s\phi _{E}\right\vert }}{2^{-\left\vert m-s\phi _{E}\right\vert }\Gamma
^{\left( -\right) }}-\frac{\left( \sqrt{-i}k_{0}r\right) ^{\left\vert
m-s\phi _{E}\right\vert }}{2^{\left\vert m-s\phi _{E}\right\vert }\Gamma
^{\left( +\right) }}\right]   \notag \\
&+&e^{i\zeta }\left[ \frac{\left( \sqrt{i}k_{0}r\right) ^{-\left\vert
m-s\phi _{E}\right\vert }}{2^{-\left\vert m-s\phi _{E}\right\vert }\Gamma
^{\left( -\right) }}\frac{\left( \sqrt{i}k_{0}r\right) ^{\left\vert m-s\phi
_{E}\right\vert }}{2^{\left\vert m-s\phi _{E}\right\vert }\Gamma ^{\left(
+\right) }}\right] ,
\end{eqnarray}%
and $\Gamma ^{\left( \pm \right) }=\Gamma \left( 1\pm \left\vert m-s\phi
_{E}\right\vert \right) $ was defined for purposes of simplification.
Inserting (\ref{nrs}) and (\ref{dright}) in (\ref{logder}), we obtain
\begin{equation}
\lim_{a\rightarrow 0^{+}}\frac{\dot{W}_{\zeta }(r)}{W_{\zeta }(r)}\Big|%
_{r=a}=\phi _{E}-s\phi _{B},  \label{saepapprox}
\end{equation}%
which gives us the parameter $\zeta $ in terms of the physics of the
problem, i.e., the correct behavior of the wave functions when $r\rightarrow
0$.

As promised above, let us now derive determine the bound states for $H_{0}$.
In order systems have a bound state, its energy must be negative, so that in
Eq. (\ref{ideal}), $k$ is a pure imaginary, i.e., $k=i\kappa $, with $\kappa
=\sqrt{-2ME}$, where $E<0$ is the bound state energy. Then, with the
substitution $k\rightarrow i\kappa $, we have
\begin{equation}
\left\{ \frac{1}{r}\frac{d}{dr}\left( r\frac{d}{dr}\right) -\left[ \frac{%
\left( m-s\phi _{E}\right) ^{2}}{r^{2}}+\kappa ^{2}\right] \right\} f_{\zeta
}(r)=0,  \label{eigenvalue}
\end{equation}%
The above equation is the modified Bessel equation whose general solution is
given by
\begin{equation}
f_{\zeta }(r)=K_{\left\vert m-s\phi _{E}\right\vert }\left( r\sqrt{-2ME}%
\right) .  \label{sver}
\end{equation}%
Since these solutions belong to $\mathcal{D}(H_{\zeta ,0})$, it is of the
form (\ref{domain}) for some $\zeta $ selected from the physics of the
problem. So, we substitute (\ref{sver}) into (\ref{domain}) and use (\ref%
{besselasympt}) to calculate the left-hand side of Eq. (\ref{logder}). After
these manipulations, we find the relation%
\begin{eqnarray}
&&\frac{\left\vert m-s\phi _{E}\right\vert \left[ a^{2\left\vert m-s\phi
_{E}\right\vert }\Gamma ^{(-)}(-ME_{b})^{\left\vert m-s\phi _{E}\right\vert
}+2^{\left\vert m-s\phi _{E}\right\vert }\Gamma ^{(+)}\right] }{%
a^{2\left\vert m-s\phi _{E}\right\vert }\Gamma ^{(-)}(-ME_{b})^{\left\vert
m-s\phi _{E}\right\vert }-2^{\left\vert m-s\phi _{E}\right\vert }\Gamma
^{(+)}}  \notag \\
&&=\phi _{E}-s\phi _{B}.  \label{derfe}
\end{eqnarray}%
Solving the above equation for $E$, we find the energy spectrum
\begin{eqnarray}
&&E=-\frac{2}{Ma^{2}}  \notag \\
&&\times \left[ \left( \frac{\phi _{E}-s\phi _{B}+\left\vert m-s\phi
_{E}\right\vert }{\phi _{E}-s\phi _{B}-\left\vert m-s\phi _{E}\right\vert }%
\right) \frac{\Gamma \left( 1+\left\vert m-s\phi _{E}\right\vert \right) }{%
\Gamma \left( 1-\left\vert m-s\phi _{E}\right\vert \right) }\right] ^{\frac{1%
}{\left\vert m-s\phi _{E}\right\vert }}.  \label{enyks}
\end{eqnarray}%
Notice that there is no arbitrary parameter in the above equation. Also, to
ensure that the energy is a real quantity, we must establish that
\begin{equation}
\left( \frac{\phi _{E}-s\phi _{B}+\left\vert m-s\phi _{E}\right\vert }{\phi
_{E}-s\phi _{B}-\left\vert m-s\phi _{E}\right\vert }\right) \frac{\Gamma
(1+\left\vert m-s\phi _{E}\right\vert )}{\Gamma (1-\left\vert m-s\phi
_{E}\right\vert )}>0.
\end{equation}%
This inequality is satisfied if%
\begin{equation}
\left\vert \phi _{E}-s\phi _{B}\right\vert \geq \left\vert m-s\phi
_{E}\right\vert .
\end{equation}%
Because of the condition that $\left\vert m-s\phi _{E}\right\vert <1$, it is
sufficient to consider $\left\vert \phi _{E}-s\phi _{B}\right\vert \geq 1$.
A necessary condition for a $\delta $ function to generate an attractive
potential, which is able to support bound states, is that the coupling
constant $\left( \phi _{E}-s\phi _{B}\right) $ must be negative. Thus, the existence of bound states requires
\begin{equation}
\phi _{E}-s\phi _{B}\leq -1.
\end{equation}%
So, it seems that we must have
\begin{equation}
s\phi _{B}>\phi _{E},
\end{equation}%
in such way that the flux and the spin must be parallel, and
consequently, a minimum value for $|\phi _{B}|$  and $|\phi _{E}|$ is established.

\section{Conclusions}

We have analyzed the planar quantum dynamics of a magnetic dipole moment in
the presence of electric and magnetic fields. We have shown that the initial
Hamiltonian system (Eq. (\ref{hm})) reduces to a planar form (Eq. (\ref{hE}%
)) without making any approximations. As an application, we have considered
the bound state problem for the case of a magnetic dipole moment interacting
with electric and magnetic fields generated by an infinitely long solenoid,
carrying a magnetic field, and with a charge density distributed uniformly
about it along the $z$-axis. The self-adjoint extension approach was used to
determine the bound states of the particle in terms of the physics of the
problem, in a very consistent way and without any arbitrary parameter.

\section{Acknowledgments}

The author would like to thank R. Casana and F. M. Andrade for the critical reading of the manuscript and for helpful discussions. This work was supported by CNPq (Grants No. 482015/2013-6 (Universal), No.306068/2013-3 (PQ)) and FAPEMA (Grant No. 00845/13 (Universal)).

\bibliographystyle{spphys}
\bibliography{dipole-v-rec.bbl}

\end{document}